\documentclass[conf]{new-aiaa}
\usepackage[utf8]{inputenc}

\usepackage{graphicx}
\usepackage{amsmath}
\usepackage[version=4]{mhchem}
\usepackage{siunitx}
\usepackage{longtable,tabularx}
\usepackage{hyperref}
\usepackage{subcaption}
\usepackage{flushend}
\usepackage{multirow}
\usepackage{url}
\title{Decentralized Coordination of Autonomous Traffic Through Advanced Air Mobility Corridors}

\author{Jasmine Jerry Aloor\footnote{Ph.D. Candidate, Department of Aeronautics and Astronautics, and AIAA Student Member.} and Hamsa Balakrishnan\footnote{ Associate Dean of Engineering, William E. Leonhard (1940) Professor, Department of Aeronautics and Astronautics, and AIAA Fellow.}}
\affil{Massachusetts Institute of Technology, Cambridge, MA 02139, USA}

\begin{document}

\maketitle

\begin{abstract}
The use of dedicated corridors for Advanced Air Mobility (AAM) traffic is one of the most commonly proposed pathways to integrating them into existing airspace operations. Most prior research has focused on the design of networks of AAM corridors and conflict resolution for aircraft within corridors. It is also generally believed that while attractive from an implementation perspective, corridor-based operations may be inefficient, especially in the absence of centralized traffic management.

In this paper, we show that contrary to this belief, it is possible for autonomous aircraft to learn to self-organize into corridor flows in decentralized settings. We illustrate our approach using scenarios in which fixed-wing aircraft need to safely and efficiently traverse (1) a single corridor with metering after the exit, (2) a sequence of two consecutive corridors, and (3) a corridor that splits into two. We find that in decentralized settings with only local information, the aircraft are able to conform to the corridor boundaries more than 94\% of the time and reach their goal in a relatively efficient manner. Furthermore, tactical interventions to handle violations of the separation minimum are needed only infrequently in low- and medium-density settings. However, such tactical interventions become more frequently necessary only when traffic density is high.

\end{abstract}

\section{Nomenclature}

{\renewcommand\arraystretch{1.0}
\noindent\begin{longtable*}{@{}l @{\quad=\quad} l@{}}
$N$ & number of agents \\
$D$ & dimension of the state \\
$S$ & state space of the environment \\
$p$ & position of agents \\
$\theta$ & heading of agent \\
$v$ & speed of agent \\
$\omega$ & angular velocity of agent \\
$a$ & longitudinal acceleration \\
$O$ & neighborhood observations \\
$\mathcal{A}$ & action space for agents \\
$\mathcal{G}$ & graph network formed by the entities in the environment  \\
$P(s'|s, A)$ & transition probability from $s$ to $s'$ given the joint action $A$ \\
$\mathcal{R}$ & joint reward function \\
$C$ & penalty applied \\
$\gamma$& discount factor \\

\end{longtable*}}

\section{Introduction\label{sec-intro}}
\lettrine{C}{orridor-based} operations are widely regarded as one of the most promising approaches to integrating Advanced Air Mobility (AAM) traffic into existing airspace systems \cite{faaUamConops,Eurocontrol-Uspace-Conops,UAE-AAM-corridor-2025,WEF-India-AAM-Conops-2024,Malawi-drone-corridor}. Existing air traffic management procedures cannot scale as the number of AAM operations grows, motivating the development of dedicated air corridors equipped with technologies that can support the new entrants \cite{FAA-AAM-implementation-plan}. The increased adoption of the AAM corridor concept has motivated work on the design, analysis, and even logistics of corridor-based AAM operations \cite{Varma-DASC2022, Tony-ICUAS2021-corridrone, He-TRE2025-air-corridor-planning}. 

The development of more automated approaches to air traffic management has been a long-standing area of research due to the growth in air traffic demand over the past few decades and the resulting increase in ATC workload \cite{TOMLIN1996,Cruck-ACC2007}. Recently, the increase in onboard sensing and computing resources, the expectation that AAM aircraft will be highly autonomous, and advances in artificial intelligence and machine learning (AI/ML) technologies have driven the study of learning-based methods for AAM traffic management. In particular, the challenge of operating multiple autonomous aircraft in the airspace has led to the consideration of Multi-agent Reinforcement Learning (MARL) as a solution methodology in the AAM context. Prior work has used MARL approaches for inter-aircraft conflict resolution \cite{isufaj2022toward,brittain2021one,ghosh2021deep}, trajectory tracking \cite{buelta2023towards}, and demand-capacity balancing \cite{spatharis-demand-capacity-balancing-2018}. 

The projected increase in demand has also motivated the development of strategic air traffic management algorithms to improve the efficiency of operations. While the desire to provide more flexibility to operators has motivated decentralized or distributed solution approaches, the approaches have largely relied on the optimization of less-constrained 4D trajectory-based operations \cite{de2021decentralized, Balakrishnan_atm2017}. The rationale behind these works has been that the constraints imposed on trajectories by AAM corridors can lead to very inefficient use of scarce airspace resources \cite{de2021decentralized}. However, a potential way to improve the efficiency of an aircraft flow through a corridor is through convoy formation, where vehicles collectively maintain an optimal separation with the vehicles in front of and behind them, as they traverse the corridor. Convoy formation or platooning has been studied in the context of railways \cite{Henke2006}, roadways \cite{Heinovski2018, Johansson2022}, and even AAM \cite{Ishihara2021convoys}, because of its potential to considerably reduce operator workload \cite{McKinsey2018} and fuel \cite{Tsugawa2016}.

The long-term vision is that Air Navigation Service Providers (e.g., the FAA) will not be responsible for providing air traffic control (ATC) services within AAM corridors \cite{faaUamConops}. This vision raises the question of whether multiple aircraft can autonomously navigate through dedicated corridor structures simultaneously in a safe and efficient manner, without centralized coordination. To the best of our knowledge, there has been very limited work on the decentralized coordination of traffic through AAM corridors. In \cite{Liu-JATM2025-Corridor-self-separation}, the authors proposed a Multi-Attribute Decision-Making scheme for aircraft to maintain self-separation within a corridor. A Model Predictive Control (MPC) based centralized approach to structured airspace networks was proposed in \cite{chour2024analysis}. By contrast, \cite{Deniz-GEIT2024-RL-structured-AAM} and \cite{Doole-aerospace2022-merge-assist} considered the problems of coordination in merges and intersections, using MARL and a merge-assist strategy based on road traffic, respectively. Finally, very recent work has developed a hybrid transformer-based MARL architecture for traffic coordination within air corridors \cite{Yu-HyTran-2025}; however, this work assumes simplified eVTOL dynamics (albeit in 3D) and that the aircraft within a corridor traverse parallel to each other. In other words, inter-aircraft separation of the traffic passing through the corridor is not considered.

Recently, we proposed InforMARL, a MARL architecture based on Graph Neural Networks (GNNs) that enables multi-agent navigation and control in limited-information settings \cite{informarl_icml}. We also extended this approach to show that agents could learn to balance fairness (load balancing between agents) and efficiency in a number of contexts, including task coverage and formation \cite{fairmarl2024}. In this paper, we address the following question: {\bf Can autonomous fixed-wing aircraft \emph{learn} to self-organize in a decentralized manner while traversing through AAM corridors?} 

\begin{figure}[htb]\centering
  \begin{subfigure}{0.33\columnwidth}
  \includegraphics[width=\textwidth]{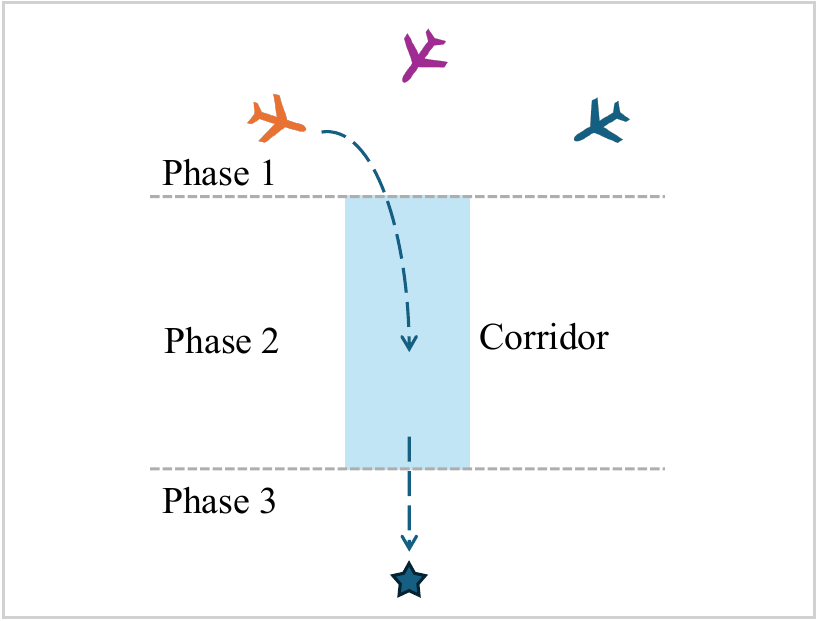}
  \caption{Single corridor with post-exit metering point}
  \label{fig:example_single}
  \end{subfigure}
  \begin{subfigure}{0.33\columnwidth}
  \includegraphics[width=\textwidth]{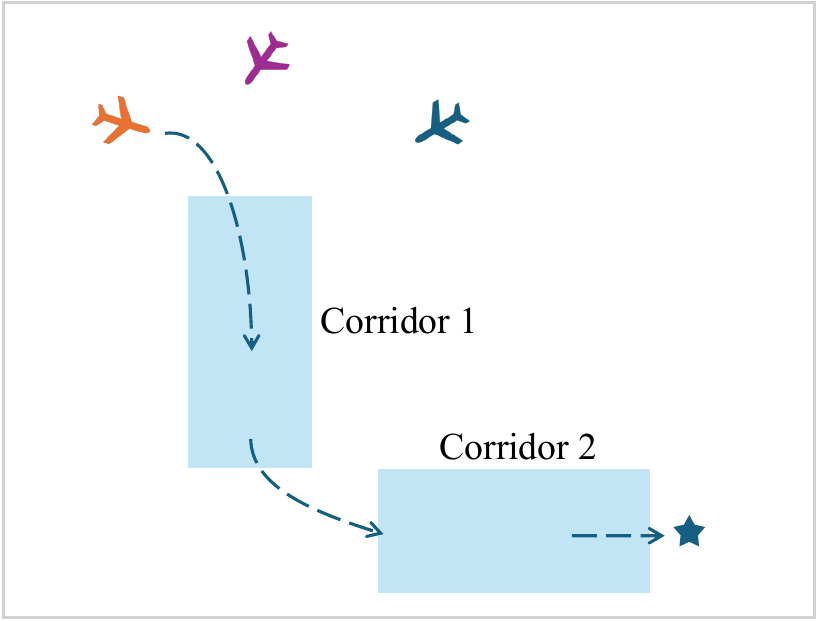}
  \caption{Two consecutive corridors placed one after another.} 
  \label{fig:example_seq}
  \end{subfigure} 
  \begin{subfigure}{0.33\columnwidth} 
  \includegraphics[width=\textwidth]{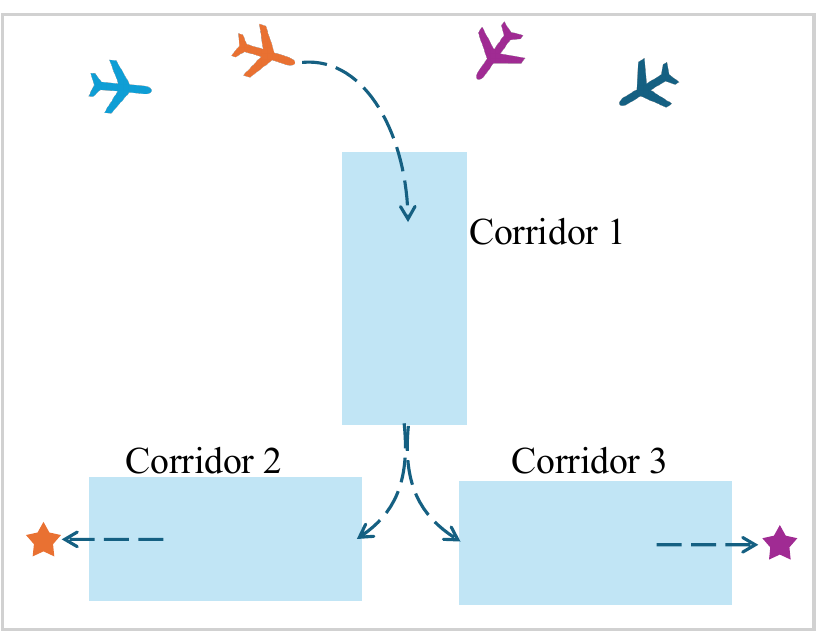} 
  \caption{Split corridor, aircraft entering a single corridor then splitting into two streams.} 
  \label{fig:example_split}
  \end{subfigure}  
  \caption{Illustrations of corridor-based scenarios\label{fig:experiments-example}.}
  \end{figure}

We show that {\bf the answer to this question is yes}, by developing a centralized training/decentralized execution approach based on our prior work. We consider three canonical scenarios, shown in Fig. \ref{fig:experiments-example}:  (a) A single corridor with a post-exit metering point that all aircraft must enter, traverse and exit; (b) two consecutive corridors, where the aircraft must enter then exit each one in succession; and (c) a split, where the traffic enters and travels through a corridor (Corridor 1 in Fig. \ref{fig:example_split}), exits it and then splits, with some aircraft entering Corridor 2 and the rest, Corridor 3. We choose these configurations as they form the building blocks of general networks of air corridors. 

Our simulations show that in all three corridor configurations, aircraft are able to conform to corridor boundaries more than $94\%$ of the time and reach their final goals in a timely manner, even in highly-congested environments. As our approach does not set hard constraints on inter-aircraft separation minima, tactical interventions may be needed when two aircraft approach within this specified value. Our experiments show that such tactical interventions are needed less than $8\%$ of the time in low- and medium-density environments, and about $17\%$ of the time in highly-congested environments with split corridors. 

\subsection{Outline}
The remainder of this paper is organized as follows. Sec. \ref{sec:problem_formulation} presents the models of agents, aircraft kinematics, and environment. Sec. \ref{sec:methodology} describes the MARL training setup, including the reward functions used. Sec. \ref{sec:results} presents our experiments and discusses the key results, and we conclude in Sec. \ref{sec:conclusions}.

\section{Models of environment and agents}
\label{sec:problem_formulation}

In this section, we describe the environment and dynamics used in our MARL training setup.

\subsection{Preliminaries}

Our environment consists of aircraft modeled as agents, goals, and obstacles, following the InforMARL \cite{informarl_icml} framework.

We model our system as a Decentralized Partially Observable Markov Decision Process (Dec-POMDP) defined by the tuple $\langle N, S, O, \mathcal{A}, \mathcal{G}, P, R,\gamma \rangle$, where:
\begin{itemize}
    \item $N$ is the number of agents
    \item $s\in S = \mathbb{R}^{N\times D}$ is the state space of the environment, with $D$ as the dimension of the state
    \item $o^{(i)}=O(s^{(i)})\in \mathbb{R}^{d}$ is the neighborhood observation for agent $i$ 
    \item $a^{(i)} \in \mathcal{A}$ is the action space for agent $i$.
    \item $g^{(i)} \in \mathcal{G}(s; i)$ is the graph network formed by the entities in the environment with respect to agent $i$
    \item $P(s'|s, A)$ is the transition probability from $s$ to $s'$ given the joint action $A$
    \item $R(s, A)$ is the joint reward function
    \item $\gamma \in [0,1)$ is the discount factor
\end{itemize}
The objective is to find a policy $\Pi= \left(\pi^{(1)}, \cdots, \pi^{(N)}\right)$, where $\pi_{\theta}^{(i)}\left(a^{(i)}|o^{(i)}, g^{(i)}\right)$ for agent $i$ selects action $a^{(i)}$ based on its graph network $g^{(i)}$ and the neighborhood observation $o^{(i)}$. 
\subsubsection{Agent dynamics}
We represent the aircraft dynamics using the mathematical model of a fixed-wing aircraft:
\begin{equation}
\begin{aligned}
 \dot{x}=v \cos \theta,\;\; \dot{y}=v \sin \theta, \;\; \dot{\theta}= \omega, \;\;\dot{v} = \mathrm{a},
\end{aligned}
\label{eq:unicycle}
\end{equation}
We define the agent state as $s^{(i)} = [x;y; \theta, v]$, which are the x- and y-positions, heading, and speed of the agent, respectively. The action space represents the angular velocity and longitudinal acceleration $a^{(i)} = [\omega, \mathrm{a}]$. Each agent's speed is limited to the range of $[v_{\min}, v_{\max}]$. Here, $v_{\max} > v_{\min} > 0$, and the agents cannot move in reverse. The action space is defined as $\mathcal{A} = [-\omega_{\max},\omega_{\max}]\times [\mathrm{a}_{\min}, \mathrm{a}_{\max}]$.
When an agent reaches a goal, it is marked as `done', and the agent is assumed to have stopped. When all agents arrive at their goals and are `done', we end the episode and start the next one.
\subsection{Corridor discretization phases}

We denote the corridors by rectangular lanes in the environment with length $L$, width $w$, and alignment angle $\theta^\textrm{corridor}$. We divide the corridor navigation process into three phases for ease of representation.

\begin{enumerate}
    \item \textbf{Phase 1}: Pre-corridor phase. Agents approach the corridor while maintaining inter-agent spacing and merge as needed to enter the corridor.
    \item \textbf{Phase 2}: In-corridor phase. Agents maintain the line formation and spacing within the corridor.
    \item \textbf{Phase 3}: Post-corridor phase. Agents exit and disperse toward their respective goals.
\end{enumerate}



\section{Methodology}
\label{sec:methodology}
We investigate a structured environment MARL navigation problem. The agents are tasked with navigating from one end of the environment to the other end. There exist corridors in the middle of the environment that act as a channel for the agents to use while navigating from one half of the environment to the other. The agents must traverse these corridors while maintaining a desired inter-agent separation. Once they cross the corridor, they have to arrive at the goal with a specified metering rate. An example is shown in Fig. \ref{fig:example_single}. 
We train our models using three agents per episode.

\subsection{Agent observations}

Each agent's local observation vector $o^{(i)}$ captures essential information about its own state, nearby agents, goals, and its designated corridor. The agent's heading, $\theta$, and speed, $v$, are recorded in a global frame of reference. The relative positions of the assigned goal and the two nearest neighboring agents are provided in the agent's body-fixed reference frame, which is aligned with its heading direction. We incorporate the details of the corridor, such as the agent's normalized longitudinal distance from the entrance $d^{\textrm{entrance}}/L$ and exit $d^{\textrm{exit}}/L$, the lateral offset from the centerline of the corridor $d^{\textrm{center}}/w$, the alignment of agent's heading to the corridor $\Delta\theta = \theta^\textrm{agent}-\theta^{\textrm{corridor}}$, and the agent’s current phase $\phi \in \{1,2,3\}$ within the corridor.
The final ego observation vector $o^{(i)}$ can be represented as $o^{(i)} = [{\theta_i}, {v_i}, p^{\textrm{goal}_1}_i, p^{\textrm{n}_1}_i, p^{\textrm{n}_2}_i, d^{\textrm{entrance}}_i/L, d^{\textrm{exit}}_i/
L, d^{\textrm{center}}_i/w, \Delta\theta, \phi_i  ]$.

Each agent also has its neighborhood information aggregated into a graph observation vector $x_j$, which is then processed by a graph neural network (GNN). Through graph message passing, the GNN can encode and incorporate information about agents and goals that lie beyond the agent's direct sensing range, and allows the framework to scale to any number of agents. For each agent $i$, $x^{(i)}_j=[p^j_i, v^j_i, p^{\mathrm{goal}_1,j}_i, \texttt{entity\_type(j)}]$ where $p^j_i, v^j_i, p^{\mathrm{goal}_1,j}_i$ are the \emph{relative} position, velocity, and position of the nearest goal of the entity at node $j$ with respect to agent $i$, respectively. The variable \texttt{entity\_type(j)} $\in \{ \texttt{agent}, \texttt{goal}\}$ specifies the type of entity at node $j$. When an agent is marked as `done', we remove the agent from the graph to prevent it from being considered in any potential conflict calculations. 

The combined input of the ego observation vector and the GNN-encoded output of neighborhood entities is provided to the policy to produce an action.

\subsection{Reward function design}
During training time, each agent is given a reward at every time step.
The reward function is designed based on the following desired behaviors.
\subsubsection{Separation maintenance}
\textit{Reward for maintaining proper inter-agent spacing}. To encourage agents to maintain an adequate inter-agent distance, we measure how close an agent is to its nearest neighbors located in front and behind it and compare it to the desired separation value ($d_s$). Any inter-agent distance that is lower than this desired spacing value is penalized.
\begin{equation}
    \mathcal{R}_\text{separation} = \sum_{\text{j} \in \{\text{front}, \text{back}\}} \max(0, d_s - \|p^{(i)} - p^{(j)}\|)
\end{equation}
\subsubsection{Phase transitions}
\textit{Reward encouraging agents to smoothly move between phases.} At the start of the episode, each agent is in Phase 1 and is provided with information about its relative position with respect to the entrance of its current corridor. The agent is provided with a continuous reward that is based on the distance to the corridor's entrance, which incentivizes them to move to the corridor. A small reward is added when the agent is near the corridor entrance to encourage heading angle alignment. When the agent transitions into Phase 2, we provide a bonus reward for completing the transition and shift the reward computation reference to the exit of the corridor. This incentivizes the agent to move toward the exit of the corridor. We provide a similar bonus phase transition reward once an agent crosses the corridor. Once an agent is in Phase 2, we provide it with a goal-reaching reward that is proportional to the distance to the agent's chosen goal. At each phase $\phi$ the reward can be computed as
\begin{equation}
    \mathcal{R}_{\mathrm{phase} (\phi)} = - \|p^\text{corridor} - p^{(i)}\| + \kappa \mathcal{R}_\text{transition}\\
\end{equation}
$\kappa$ is set to 1 if the agent made the phase transition for the first time.
To prevent agents from skipping the correct phase transition order or trying to move in the reverse phase order, we penalize the agent with an incorrect phase transition penalty $-C_\textrm{phase}$.

\subsubsection{Conflict avoidance}
\textit{Penalty for approaching too close to other agents or obstacles.} We also penalize agents colliding with other agents in the environment using a conflict penalty $-C_\textrm{conflict}$. This penalty is applied when the inter-agent separation becomes lower than the desired minimum separation distance $d_s$.

\subsubsection{Goal reaching reward}
\textit{Reward encouraging a quick approach to the goal.} When an agent successfully reaches its goal (indicated by $\rho$), it receives a one-time goal-reaching reward $\mathcal{R}_\mathrm{goal}$. The indicator variable $\rho$ is 1 if the agent reached the assigned goal and was previously not at the goal; otherwise, 0.

The overall reward provided to each agent at every time step then becomes,
\begin{equation}
    \mathcal{R}_{\mathrm{total}} = \mathcal{R}_\text{separation} +\mathcal{R}_{\mathrm{phase} (\phi)}   +\rho \mathcal{R}_\mathrm{goal} -C_\textrm{phase} -C_\textrm{conflict}
\end{equation}
This reward design ensures that the agents learn to maintain a desired separation while being able to follow the various phases of the corridor and reach the desired goal.
\section{Experiments and results}
\label{sec:results}
\subsection{Parameter values}
We consider the following parameters for the environment and aircraft kinematics. The parameter values are chosen to broadly align with industry standards and proposed corridor designs \cite{wisk,joby,uam_separation, AloorJ-RSS-25}. We note that these particular parameter values are chosen to merely demonstrate our methodology and can be changed as technology and standards evolve. Additionally, the minimum inter-aircraft separation could be adjusted to a fuel-optimal (or some other preferred) separation to achieve convoy formation in high-demand scenarios.  
\begin{enumerate}
    \item Minimum groundspeed, $v_{\text{min}}$: 60 knots/s = 111 km/h
    \item Maximum groundspeed, $v_{\text{max}}$: 175 knots/s = 324 km/h
    \item Minimum acceleration, $a_{\text{min}}$: $-1$ m/s$^2$
    \item Maximum acceleration, $a_{\text{max}}$: 2 m/s$^2$
    \item Maximum angular velocity, $\omega_{\max}$: 0.075 rad/s
    \item Length of air corridor: between 2 km to 1.2 km
    \item Width of air corridor: 0.6 km
    \item Minimum inter-aircraft separation: 0.3 km
    \item Goal threshold distance: 0.2 km
    \item Metering rate at goal: 10 aircraft/min
    \item Episode length: 150 s
\end{enumerate}
\subsection{Experimental setup}
To evaluate our approach, we define the following test scenarios, each designed to assess different aspects of corridor-based aircraft navigation. 

\subsubsection{Single corridor with post-exit metering point}
In this scenario, illustrated in Fig. \ref{fig:example_single}, aircraft follow a structured, metered entry into a single corridor. The objective is to evaluate how well the method regulates the flow of aircraft, ensuring smooth entry and maintaining safe separation between them. This is the simplest scenario and serves as a baseline to verify that autonomously operated aircraft can traverse a corridor and satisfy metering requirements at a point downstream.

For each episode, we define the air corridor to be at the center of the environment. We vary the initial position of each aircraft for better generalization. Each aircraft is required to maintain the desired inter-aircraft separation minimum relative to all other aircraft. The vehicles are tasked to go to the goal at a fixed metering rate. Once the aircraft reaches its goal, it is stopped and not considered for any further actions for that episode.

\begin{figure}
    \centering
    \includegraphics[width=0.3\linewidth]{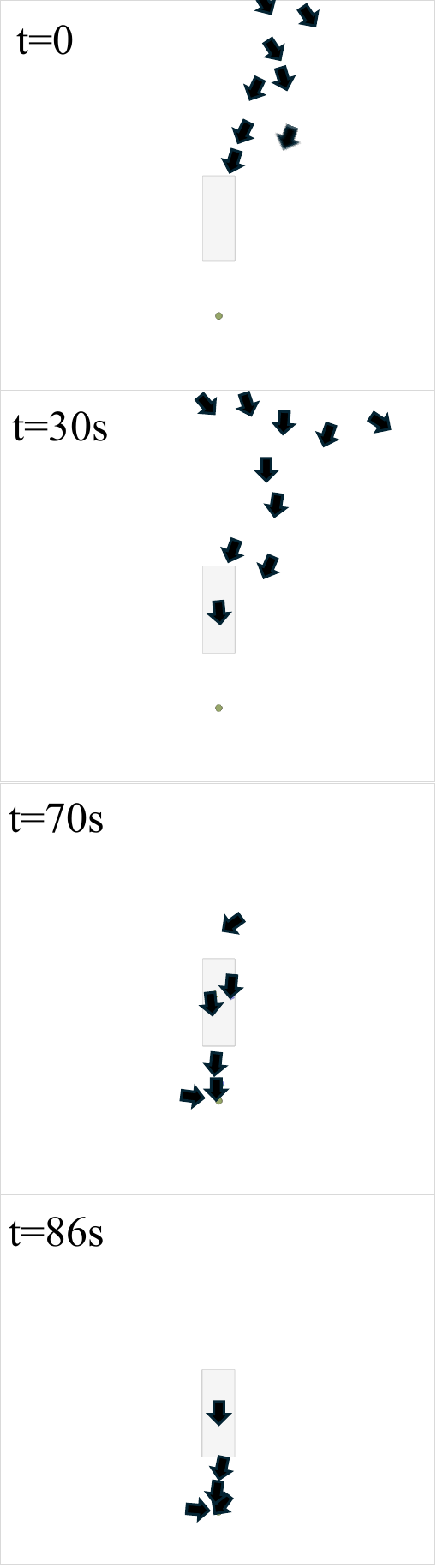}
    \caption{Visualization of behaviors of 10 aircraft in the simple metered corridor scenario. The aircraft start from the upper half of the environment and are tasked to navigate through the corridor and reach a point. The length of the vertical corridor is 1.6 km. The arrows indicate aircraft positions and are not to scale.}
    \label{fig:simple_metered_experiment}
\end{figure}
\subsubsection{Two consecutive corridors}
This scenario, as shown in Fig. \ref{fig:example_seq}, increases complexity by requiring agents to traverse multiple consecutive corridors. The agents are tasked with going through Corridor 1 first before entering the horizontal Corridor 2. 

In each episode, each aircraft is provided with information about Corridor 1 via its local observations. Once the aircraft traverses the first corridor, we provide information about the second corridor. This allows the multi-agent system to be scaled up to multiple corridors through a simple repetition of the corridor setup. Once the aircraft crosses the second corridor, it proceeds to its destination and comes to a stop there.
\subsubsection{Split corridor}
This scenario, depicted in Fig. \ref{fig:example_split}, adds to the previous scenario by dynamically allocating a second corridor to each aircraft once it traverses Corridor 1. This scenario demonstrates the ability of our method to accommodate forks in the route.
In this scenario, every aircraft has a common goal of traversing through Corridor 1. As it exits the corridor, each aircraft is randomly assigned to either Corridor 2 or Corridor 3. The aircraft stops when it reaches the goal at the end of its assigned corridor.


\begin{figure}
    \centering
    \includegraphics[width=0.3\linewidth]{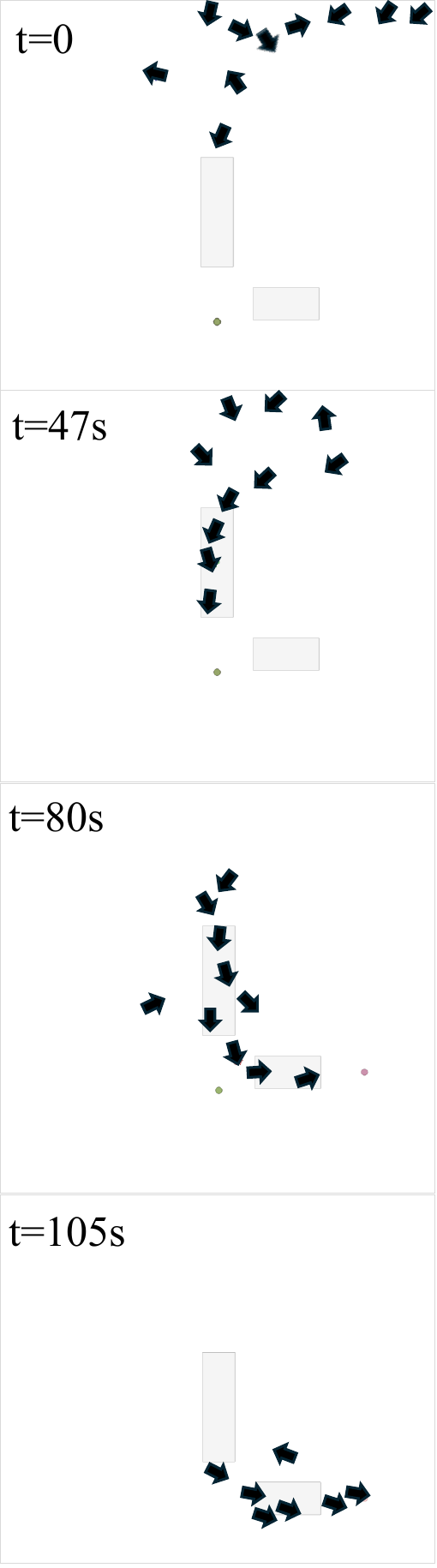}
    \caption{Visualization of behaviors of 10 aircraft with the two consecutive corridors scenario. The aircraft start from the upper half of the environment and are tasked to navigate through both corridors. The length of the first vertical corridor is 2.0 km. The length of the horizontal corridor is 1.2 km. The arrows indicate aircraft positions and are not to scale.}
    \label{fig:seq-corridor_experiment}
\end{figure}

\begin{figure}
    \centering
    \includegraphics[width=0.3\linewidth]{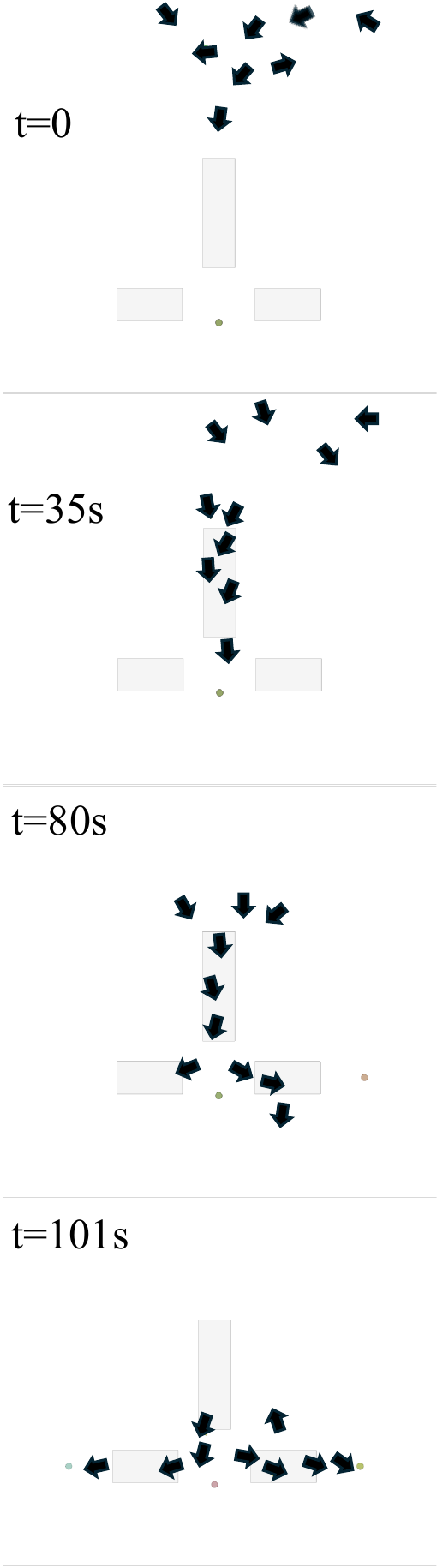}
    \caption{Visualization of behaviors of 10 aircraft in the split scenario where all aircraft enter the first corridor and then split into two corridors. The length of the first vertical corridor is 2.0 km. The length of each horizontal corridor is 1.2 km. The arrows indicate aircraft positions and are not to scale.}
    \label{fig:split-corridor}
\end{figure}

\subsection{Performance metrics}

We run evaluations for each scenario for 100 randomly generated episodes. For each scenario, we consider traffic flows with 3, 5, and 10 aircraft. We note that the corridors are either 1.6 or 3.2 km in length; consequently, the aircraft counts correspond to low, medium, and very high density operations. We calculate the following performance metrics for each scenario in our experiments.

\subsubsection{Conformance to corridor boundaries ($C\%$)}
This metric measures how well aircraft adhere to the designated corridor boundaries during their traversal. Conformance is evaluated based on the deviation of an aircraft’s trajectory from the corridor boundaries, ensuring the aircraft remains within the allowable corridor width. $C\%$ is calculated as the average percentage of time that each aircraft stays within the corridor boundaries while traversing through it, averaged over all aircraft and 100 episodes. High conformance to corridor boundaries is equivalent to the required navigation performance and reflects the effectiveness of the AAM corridor definition.

\subsubsection{Success rate ($S\%$)}
Success rate represents the percentage of aircraft that successfully pass through the corridors and reach their assigned goal within the length of the episode, i.e., 150 s. A high value of the success rate indicates better performance, as it implies that the aircraft flies efficient routes.

\subsubsection{Completion time per episode ($T$, s)}
The time taken per episode is calculated based on how long it takes for all the aircraft in a scenario to reach their goals. A lower value of $T$ is reflective of efficient navigation by all aircraft without unnecessary delays.

\subsubsection{Violation of separation minimum ($\Delta d$, m)}
$\Delta d$ is calculated as the amount by which the inter-aircraft separation minimum, which is assumed to be 300 m, is violated. It is only considered when two aircraft come closer than the separation minimum; in these situations, $\Delta d$ is given by the difference between the minimum and actual separations. For example, if two aircraft are 290 m apart, then $\Delta d = 10$ m. This metric quantifies how well aircraft satisfy the separation minima while traveling through corridors. We report the mean and standard deviation of $\Delta d$ in meters, noting that the values are only averaged over instances when a violation occurs. A lower value of $\Delta d$ is preferred. We note that in all our experiments, the inter-aircraft separation never became zero, i.e., there were no collisions even without tactical deconfliction.

\subsubsection{Need for tactical intervention for deconfliction ($I\%$)}
In practice, a tactical intervention will be activated any time that two aircraft approach each other closer than the separation minimum. We therefore also calculate the need for such tactical intervention, $I$, as the average fraction of time when aircraft within corridors have a separation violation relative to the total time they spend in corridors. As in the case of $\Delta d$, a lower value of $I$ is preferred.


Table \ref{table_results} shows these metrics for the three corridor scenarios described in Fig. \ref{fig:experiments-example}.
\begin{table}[ht!]
\caption{Performance metrics for the three corridor scenarios illustrated in Fig. \ref{fig:experiments-example}. The metrics presented are the conformance to corridor boundaries ($C\%$, higher is better), the success rate ($S\%$, higher is better), the average completion time $T$ per episode in seconds (lower is better), the mean and standard deviation of the separation distances when the minimum separation is violated ($\Delta d$ m, lower is better), and the fraction of time that a tactical intervention for deconfliction is needed ($I\%$, lower is better). These performance metrics are defined in Sec. \ref{sec:results} C. We consider environments with 3, 5, and 10 aircraft, representing low-, medium-, and high-density environments.}
\centering
\begin{tabular}{|c||c||ccccc|}
\hline
\multirow{2}{*}{\textbf{Scenario}} & \multicolumn{1}{c||}{\multirow{2}{*}{\textbf{\# agents}}} & \multicolumn{1}{c|}{Conformance, $C$} & \multicolumn{1}{c|}{Success, $S$} & \multicolumn{1}{c|}{Completion, $T$} & \multicolumn{1}{c|}{$\Delta d$} & \multicolumn{1}{c|}{Need for Tactical} \\
 & & \multicolumn{1}{c|}{(\%)} & \multicolumn{1}{c|}{(\%)} & \multicolumn{1}{c|}{(s)} & \multicolumn{1}{c|}{(m)} & \multicolumn{1}{c|}{Deconfliction, $I$ (\%)} \\ 
\hline \hline
\multirow{3}{*}{Single} & 3 & \multicolumn{1}{c|}{99\%} & \multicolumn{1}{c|}{99\%} & \multicolumn{1}{c|}{60.5} & \multicolumn{1}{c|}{0±36} & 1\%  \\ \cline{2-7} 
 & 5 & \multicolumn{1}{c|}{95\%} & \multicolumn{1}{c|}{94\%} & \multicolumn{1}{c|}{63.9} & \multicolumn{1}{c|}{1±52} & 8\% \\ \cline{2-7} 
 & 10 & \multicolumn{1}{c|}{94\%} & \multicolumn{1}{c|}{92\%} & \multicolumn{1}{c|}{86.1} & \multicolumn{1}{c|}{97±50} & 14\%  \\ \hline \hline

\multirow{3}{*}{Consecutive} & 3 & \multicolumn{1}{c|}{98\%} & \multicolumn{1}{c|}{98\%} & \multicolumn{1}{c|}{96.8} & \multicolumn{1}{c|}{0±41} & 1\% \\ \cline{2-7} 
 & 5 & \multicolumn{1}{c|}{95\%} & \multicolumn{1}{c|}{97\%} & \multicolumn{1}{c|}{99.9} & \multicolumn{1}{c|}{8±56} &  6\% \\ \cline{2-7} 
 & 10 & \multicolumn{1}{c|}{94\%} & \multicolumn{1}{c|}{83\%} & \multicolumn{1}{c|}{102.5} & \multicolumn{1}{c|}{97±42} & 17\%\\ \hline \hline
\multirow{3}{*}{Split} & 3 & \multicolumn{1}{c|}{95\%} & \multicolumn{1}{c|}{90\%} & \multicolumn{1}{c|}{87.8} & \multicolumn{1}{c|}{0±48} &  2\% \\ \cline{2-7} 
 & 5 & \multicolumn{1}{c|}{95\%} & \multicolumn{1}{c|}{88\%} & \multicolumn{1}{c|}{91.5} & \multicolumn{1}{c|}{31±53} & 7\%\\ \cline{2-7} 
 & 10 & \multicolumn{1}{c|}{96\%} & \multicolumn{1}{c|}{87\%} & \multicolumn{1}{c|}{109.9} & \multicolumn{1}{c|}{41±127} & 12\%\\ \hline

\end{tabular}
\label{table_results}
\end{table}

\subsection{Discussion}
In all three scenarios, aircraft maintain high conformance to the corridor even when the total number of agents in the system is modified.
The single corridor with the post-exit metering point achieves high success rates while also resulting in a lower violation of the separation minimum and the need for tactical intervention for deconfliction. Owing to the simple environment structure, it also takes the shortest total time. As we increase the number of aircraft from what was used in training to 5 and then 10 aircraft, as shown in Fig. \ref{fig:simple_metered_experiment}, we observe a decrease in the corridor conformance and increased violations of the separation minimum.

As we increase the complexity of the scenario, such as the consecutive and split corridor experiments, we observe a lower success rate. Additionally, with an increased number of aircraft, the need for tactical intervention increases due to congestion in the environment. Fig. \ref{fig:seq-corridor_experiment} shows a series of time steps of an episode involving the consecutive corridor scenario in a congested environment with ten aircraft.
With the split corridor scenario (Fig. \ref{fig:split-corridor}), the aircraft have multiple options and have to follow their recommended corridors, which further increases the complexity. This impacts the performance, as seen in the lower success rates across all numbers of agents.  


\section{Conclusions and future work}
\label{sec:conclusions}
In this work, we presented a method that enables autonomous fixed-wing aircraft to learn to self-organize while traversing through AAM corridors in a decentralized manner. We modeled three scenarios involving aircraft traveling through corridors with increasing levels of complexity. The simplest scenario involved aircraft traversing through a single corridor with a post-exit metering point. We added consecutive corridors and split corridors and performed experiments with varying numbers of aircraft to represent levels of congestion. Our results showed that aircraft are able to conform to the corridor boundaries within the specified episode time. We also showed the requirements of tactical interventions when aircraft violate the desired spacing minima.

Future work includes extending the corridor-based navigation to coverage tasks after traversing the corridor and considering a greater number of agents. Other interesting directions are generalizations to unseen environments and 3D vehicle dynamics with dynamic corridor generation.

\section*{Acknowledgments}
The authors would like to thank the MIT SuperCloud \cite{supercloud} and the Lincoln Laboratory Supercomputing Center for providing high-performance computing resources that have contributed to the research results reported within this paper. We also thank Ruhundaka Ejilemele and Aadarsh Govada for helpful discussions. J. Aloor was supported in part by a Mathworks Fellowship. This work was supported by NASA under grant \#80NSSC23M0220, but this article solely reflects the opinions and conclusions of its authors and not any NASA entity. 

\bibliography{sample}

@article{isufaj2022toward,
  title={Toward conflict resolution with deep multi-agent reinforcement learning},
  author={Isufaj, Ralvi and Aranega Sebastia, David and Angel Piera, Miquel},
  journal={Journal of Air Transportation},
  volume={30},
  number={3},
  pages={71--80},
  year={2022},
  publisher={American Institute of Aeronautics and Astronautics},
  url = {https://doi.org/10.2514/1.D0296}
}

@inproceedings{buelta2023towards,
  title={Towards multi-aircraft transfer learning for trajectory tracking},
  author={Buelta, Almudena and Olivares, Alberto and Staffetti, Ernesto},
  booktitle={Proceedings of the Fifteenth USA/Europe Air Traffic Management Research and Development Seminar},
  year={2023},
  url = {https://hdl.handle.net/10115/69757}
}

@inproceedings{ghosh2021deep,
  title={A deep ensemble method for multi-agent reinforcement learning: A case study on air traffic control},
  author={Ghosh, Supriyo and Laguna, Sean and Lim, Shiau Hong and Wynter, Laura and Poonawala, Hasan},
  booktitle={Proceedings of the International Conference on Automated Planning and Scheduling},
  volume={31},
  pages={468--476},
  year={2021},
  url = {https://doi.org/10.1609/icaps.v31i1.15993}
}

@inproceedings{brittain2021one,
  title={One to any: Distributed conflict resolution with deep multi-agent reinforcement learning and long short-term memory},
  author={Brittain, Marc W and Wei, Peng},
  booktitle={AIAA Scitech 2021 Forum},
  pages={1952},
  year={2021},
  url = {https://doi.org/10.2514/6.2021-1952}
}

@article{TOMLIN1996,
title = {Hybrid Control in Air Traffic Management Systems},
journal = {IFAC Proceedings Volumes},
volume = {29},
number = {1},
pages = {5512-5517},
year = {1996},
author = {C. Tomlin and G. Pappas and J. Lygeros and D. Godbole and S. Sastry and G. Meyer},
URL = {https://www.sciencedirect.com/science/article/pii/S1474667017585596}
}

@InProceedings{informarl_icml,
  title = 	 {Scalable Multi-Agent Reinforcement Learning through Intelligent Information Aggregation},
  author =       {Nayak, Siddharth and Choi, Kenneth and Ding, Wenqi and Dolan, Sydney and Gopalakrishnan, Karthik and Balakrishnan, Hamsa},
  booktitle = 	 {Proceedings of the 40th International Conference on Machine Learning},
  pages = 	 {25817--25833},
  year = 	 {2023},
  volume = 	 {202},
  series = 	 {Proceedings of Machine Learning Research},
  month = 	 {23--29 Jul},
  publisher =    {PMLR},
URL ={https://proceedings.mlr.press/v202/nayak23a.html}
}

@article{fairmarl2024,
author = {Aloor, Jasmine Jerry and Nayak, Siddharth Nagar and Dolan, Sydney and Balakrishnan, Hamsa},
title = {Cooperation and Fairness in Multi-Agent Reinforcement Learning},
year = {2024},
issue_date = {June 2025},
publisher = {Association for Computing Machinery},
address = {New York, NY, USA},
volume = {2},
number = {2},
url = {https://doi.org/10.1145/3702012},
doi = {10.1145/3702012},
journal = {ACM J. Auton. Transport. Syst.},
month = dec,
articleno = {8},
numpages = {25}
}

@inproceedings{de2021decentralized,
  title={Decentralized air traffic management for advanced air mobility},
  author={de Oliveira, {\'I}talo Romani and Neto, Euclides Carlos Pinto and Matsumoto, Thiago Toshio and Yu, Huafeng},
  booktitle={2021 Integrated Communications Navigation and Surveillance Conference (ICNS)},
  pages={1--8},
  year={2021},
  organization={IEEE},
  url = "https://doi.org/10.1109/ICNS52807.2021.9441552"
}

@inproceedings{spatharis-demand-capacity-balancing-2018,
author = {Spatharis, Christos and Kravaris, Theocharis and Vouros, George A. and Blekas, Konstantinos and Chalkiadakis, Georgios and Garcia, Jose Manuel Cordero and Fernandez, Esther Calvo},
title = {Multiagent Reinforcement Learning Methods to Resolve Demand Capacity Balance Problems},
year = {2018},
isbn = {9781450364331},
publisher = {Association for Computing Machinery},
address = {New York, NY, USA},
url = {https://doi.org/10.1145/3200947.3201010},
doi = {10.1145/3200947.3201010},
booktitle = {Proceedings of the 10th Hellenic Conference on Artificial Intelligence},
articleno = {2},
numpages = {9},
location = {Patras, Greece},
series = {SETN '18}
}

@inproceedings{Ishihara2021convoys,
  title={{A Framework for Sense and Follow Convoys for Collective Autonomous Mobility}},
  author={Abraham K. Ishihara and Husni R. Idris and Min Xue},
  booktitle={AIAA Aviation Forum}, 
  year={2021},
  url ={https://doi.org/10.2514/6.2021-2329}
}

@techreport{faaUamConops,
  title       = "{Urban Air Mobility (UAM) Concept of Operations Version 2.0}",
  author = {{Federal Aviation Administration}},
  institution = "Federal Aviation Administration",
  address     = "Washington, DC",
  year        = "2023",
  month       = "April",
  url = {https://www.faa.gov/air-taxis/uam_blueprint}
}

@techreport{FAA-AAM-implementation-plan,
  title       = "{Advanced Air Mobility (AAM)
Implementation Plan}",
  author = {{Federal Aviation Administration}},
  institution = "Federal Aviation Administration",
  address     = "Washington, DC",
  year        = "2023",
  month       = "July",
  url = "https://www.faa.gov/air-taxis/implementation-plan"
}

@techreport{Eurocontrol-Uspace-Conops,
  title       = "{U-space ConOps and
architecture (edition 4)}",
  author = {{SESAR Joint Undertaking}},
  institution = "EUROCONTROL",
  year        = "2023",
  month       = "July",
  url ="https://www.sesarju.eu/node/4544"
}

@article{Liu-JATM2025-Corridor-self-separation,
title = {Enhanced self-separation decision making for autonomous flight operations in air corridors},
journal = {Journal of Air Transport Management},
volume = {124},
pages = {102721},
year = {2025},
issn = {0969-6997},
doi = {https://doi.org/10.1016/j.jairtraman.2024.102721},
url = {https://www.sciencedirect.com/science/article/pii/S0969699724001868},
author = {Zhaoxuan Liu and Maolin Wang and Zhiyong Liu}
}

@article{He-TRE2025-air-corridor-planning,
title = {Air Corridor Planning for Urban Drone Delivery: Complexity Analysis and Comparison via Multi-Commodity Network Flow and Graph Search},
journal = {Transportation Research Part E: Logistics and Transportation Review},
volume = {193},
pages = {103859},
year = {2025},
issn = {1366-5545},
doi = {https://doi.org/10.1016/j.tre.2024.103859},
url = {https://www.sciencedirect.com/science/article/pii/S1366554524004502},
author = {Xinyu He and Lishuai Li and Yanfang Mo and Zhankun Sun and S. Joe Qin}
}

@article{Deniz-GEIT2024-RL-structured-AAM,
title = {A reinforcement learning approach to vehicle coordination for structured advanced air mobility},
journal = {Green Energy and Intelligent Transportation},
volume = {3},
number = {2},
pages = {100157},
year = {2024},
issn = {2773-1537},
doi = {https://doi.org/10.1016/j.geits.2024.100157},
url = {https://www.sciencedirect.com/science/article/pii/S2773153724000094},
author = {Sabrullah Deniz and Yufei Wu and Yang Shi and Zhenbo Wang}
}

@INPROCEEDINGS{Tony-ICUAS2021-corridrone,
  author={Tony, Lima Agnel and Ratnoo, Ashwini and Ghose, Debasish},
  booktitle={2021 International Conference on Unmanned Aircraft Systems (ICUAS)}, 
  title={Lane Geometry, Compliance Levels, and Adaptive Geo-fencing in CORRIDRONE Architecture for Urban Mobility}, 
  year={2021},
  volume={},
  number={},
  pages={1611-1617},
  doi={10.1109/ICUAS51884.2021.9476745}}

@Article{Doole-aerospace2022-merge-assist,
AUTHOR = {Doole, Malik and Ellerbroek, Joost and Hoekstra, Jacco M.},
TITLE = {Investigation of Merge Assist Policies to Improve Safety of Drone Traffic in a Constrained Urban Airspace},
JOURNAL = {Aerospace},
VOLUME = {9},
YEAR = {2022},
NUMBER = {3},
ARTICLE-NUMBER = {120},
URL = {https://www.mdpi.com/2226-4310/9/3/120},
ISSN = {2226-4310},
DOI = {10.3390/aerospace9030120}
}

@article{UAE-AAM-corridor-2025,
author = {Garrett Reim},
month = {February},
 year = {2025},
 title = {{UAE Begins Mapping Air Corridors For Air Taxis, Cargo Drones}},
 journal = {Aviation Week},
 url = {http://aviationweek.com/aerospace/advanced-air-mobility/uae-begins-mapping-air-corridors-air-taxis-cargo-drones},
 urldate = {2025-02-14},
}

@Online{WEF-India-AAM-Conops-2024,
 year = {2024},
author = {{World Economic Forum}},
 title = {{Skyways to the Future:
Operational Concepts for
Advanced Air Mobility in India}},
 note={\url{http://reports.weforum.org/docs/WEF_Skyways_to_the_Future_2024.pdf}},
}

@Online{Malawi-drone-corridor,
 year = {2017},
author = {{UNICEF}},
 title = {{Malawi's Unique Drone Corridor}},
 note = {\url{https://www.unicef.org/innovation/drones/malawi-unique-drone-corridor}},
 urldate = {2017-07-03},
}

@INPROCEEDINGS{Varma-DASC2022,
  author={Verma, Savvy and Dulchinos, Victoria and Wood, Robert Dan and Farrahi, Amir and Mogford, Richard and Shyr, Megan and Ghatas, Rania},
  booktitle={2022 IEEE/AIAA 41st Digital Avionics Systems Conference (DASC)}, 
  title={Design and Analysis of Corridors for UAM Operations}, 
  year={2022},
  volume={},
  number={},
  pages={1-10},
  doi={10.1109/DASC55683.2022.9925820}}

@inproceedings{supercloud,
title={Interactive supercomputing on 40,000 cores for machine learning and data analysis},
author={Reuther, Albert and Kepner, Jeremy and Byun, Chansup and Samsi, Siddharth and Arcand, William and Bestor, David and Bergeron, Bill and Gadepally, Vijay and Houle, Michael and Hubbell, Matthew and Jones, Michael and Klein, Anna and Milechin, Lauren and Mullen, Julia and Prout, Andrew and Rosa, Antonio and Yee, Charles and Michaleas, Peter},
booktitle={2018 IEEE High Performance extreme Computing Conference (HPEC)},
pages={1--6},
year={2018},
organization={IEEE},
url = {https://doi.org/10.1109/HPEC.2018.8547629}
}

@INPROCEEDINGS{Cruck-ACC2007,
  author={Cruck, Eva and Lygeros, John},
  booktitle={2007 American Control Conference}, 
  title={Subliminal air traffic control: Human friendly control of a multi-agent system}, 
  year={2007},
  volume={},
  number={},
  pages={462-467},
  doi={10.1109/ACC.2007.4282641}}

@TECHREPORT{uam_separation,
  author={ Mogford, Richard and Peknik, Dan and Zelman, Jake},
  title={{UAM} Variable Separation},
  institution={NASA Ames Research Center},
  number={},
  type={Presentation},
  year={2020},
 note = {\url{https://ntrs.nasa.gov/citations/20205006214}},
}

@inproceedings{Balakrishnan_atm2017, 
  author={Hamsa Balakrishnan and Bala Chandran},
  title={A Distributed Framework for Traffic Flow
Management in the Presence of Unmanned Aircraft},
 booktitle = {USA/Europe ATM R\&D Seminar}, 
year={2017},
url = {https://dspace.mit.edu/handle/1721.1/114703}
 }

@inproceedings{Henke2006,
author = {C. Henke and N. Frohleke and J. Bocker},
year = {2006},
title = {Advanced convoy control strategy for autonomously driven railway vehicles},
booktitle={IEEE Intelligent Transportation Systems Conference},
url = {https://doi.org/10.1109/ITSC.2006.1707417}
}

@article{Johansson2022, 
 title={Strategic Hub-Based Platoon Coordination Under Uncertain Travel Times}, 
 journal={IEEE Transactions on Intelligent Transportation Systems}, 
 author={A. Johansson and E. Nekouei and K. H. Johansson and J. Mårtensson}, 
 year={2022},
 url = "https://doi.org/10.1109/TITS.2021.3077467"
 }

@inproceedings{Heinovski2018,
author = {J. Heinovski and F. Dressler},
year = {2018},
title = {Platoon formation: Optimized car to platoon assignment strategies and protocols},
booktitle={IEEE
Vehicular Networking Conference (VNC)},
url = "https://doi.org/10.1109/VNC.2018.8628396"
}

@misc{McKinsey2018,
author = {A. Chottani and G. Hastings and J. Murnane and F. Neuhaus},
title = {{Distraction or disruption? Autonomous trucks gain ground in US logistics}},
howpublished = {\url{www.mckinsey.com/industries/travel-logistics-and-infrastructure/our-insights/distraction-or-disruption-autonomous-trucks-gain-ground-in-us-logistics}},
year = {2018},
}

@article{Tsugawa2016, 
 title={A review of truck platooning projects for energy savings}, 
 journal={IEEE Transactions on Intelligent Vehicles}, 
 author={S. Tsugawa and S. Jeschke and S. E. Shladover}, 
 year={2016},
 url ={https://ieeexplore.ieee.org/document/7497531}
 }

@misc{joby,
key = {Joby Aviation},
year = {2025},
note = {\url{https://www.jobyaviation.com/}},
}

@misc{wisk,
key = {Wisk},
year = {2025},
 note = {\url{https://wisk.aero/aircraft/}},
}

@ARTICLE{Yu-HyTran-2025,
author={Yu, Liangkun and Li, Zhirun and Ansari, Nirwan and Sun, Xiang},
journal={ IEEE Transactions on Mobile Computing},
title={{ Hybrid Transformer Based Multi-Agent Reinforcement Learning for Multiple Unmanned Aerial Vehicle Coordination in Air Corridors }},
year={2025},
volume={},
number={01},
ISSN={1558-0660},
pages={1-14},
doi={10.1109/TMC.2025.3532204},
url = {https://doi.ieeecomputersociety.org/10.1109/TMC.2025.3532204},
publisher={IEEE Computer Society},
address={Los Alamitos, CA, USA},
month=jan}

@inproceedings{chour2024analysis,
  title={Analysis of Traffic Flow in Structured Urban Airspace Networks with MFD-based Feedback Control},
  author={Chour, Kenny and Razzaghi, Pouria and Verma, Dhriti and Xue, Min and Munishkin, Alexey and Kalyanam, Krishna},
  booktitle={AIAA AVIATION FORUM AND ASCEND 2024},
  pages={4575},
  year={2024},
  url ={https://doi.org/10.2514/6.2024-4575}
}

@INPROCEEDINGS{AloorJ-RSS-25, 
    AUTHOR    = {Jason Jangho Choi AND Jasmine Jerry Aloor AND Jingqi Li AND Maria G. Mendoza AND Hamsa Balakrishnan AND Claire Tomlin}, 
    TITLE     = {{Resolving Conflicting Constraints in Multi-Agent Reinforcement Learning with Layered Safety}}, 
    BOOKTITLE = {Proceedings of Robotics: Science and Systems}, 
    YEAR      = {2025}, 
    ADDRESS   = {LosAngeles, CA, USA}, 
    MONTH     = {June}, 
    DOI       = {10.15607/RSS.2025.XXI.094} 
}

\end{document}